\documentclass[aps,prl,twocolumn,floatfix,showpacs]{revtex4}

\usepackage{graphics}
\usepackage{bm}

\def\PRA{{Phys.~Rev.~A} }

\def\PRL{{Phys.~Rev.~Lett.} }

\newcommand{\myscalebox}[1]{\scalebox{0.45}[0.45]{#1}}
\newcommand{\myscaleboxa}[1]{\scalebox{0.45}[0.3]{#1}}
\newcommand{\myscaleboxb}[1]{\scalebox{0.4}[0.4]{#1}}

\newcommand{\be}{\begin{equation}}
\newcommand{\bea}{\begin{eqnarray}}
\newcommand{\eea}{\end{eqnarray}}
\newcommand{\ee}{\end{equation}}

\begin{document}

\title{Improved Lewenstein model for high-order harmonic generation
of atoms and molecules with scattering wavefunctions}

\author{Anh-Thu Le,\footnote{Email address: {atle@phys.ksu.edu}}$^1$
Toru Morishita,$^{1,2}$ and C.~D. Lin$^1$}

\affiliation{$^1$Department of Physics, Cardwell Hall, Kansas
State University, Manhattan, KS 66506, USA\\
$^2$Department of Applied Physics and Chemistry, University of
Electro-Communications, 1-5-1 Chofu-ga-oka, Chofu-shi, Tokyo,
182-8585, Japan}

\date{\today}

\begin{abstract}

We demonstrate a simple method to improve the Lewenstein model for
the description of high-order harmonic generation (HHG). It is
shown that HHG spectra can be expressed as the product of a
returning electron wave packet and the photo-recombination cross
sections, where the former can be extracted from the Lewenstein
model. By replacing plane waves with scattering waves in the
calculation of recombination matrix elements, we showed that the
resulting HHG spectra agree well with those from solving the
time-dependent Schr\"odinger equation. The improved model can be
used for quantitative calculations of high harmonics generated by
molecules.

\end{abstract}


\maketitle

 When an atom is subjected to a strong driving laser field, one of
 the nonlinear response processes is the generation of high-order
 harmonics. In the past decade, high-order harmonic generation (HHG) has been used
 extensively for the generation of single attosecond pulses \cite{drescher,sekikawa,sansone}
 and attosecond pulse trains \cite{apt},
 thus opening up new opportunities for attosceond time-resolved
 spectroscopy. HHG is understood using the
 three-step model \cite{corkum,lewenstein} -- first the electron is released by tunnel
 ionization; second, it is accelerated by the oscillating electric
 field of the laser and later driven back to the target
 ion; and third, the electron recombines with the ion to emit
 a high energy photon. A semiclassical formulation of the three-step
 model based on the strong field approximation (SFA) is given by
 Lewenstein {\it et al} \cite{lewenstein}.
 In this SFA model (often called Lewenstein model),
 the liberated continuum electron experiences the full effect from the laser
 field, but not from the ion that it has left behind. In spite of
 this limitation, Lewenstein model has been widely used for understanding
 the HHG by atoms and molecules. Since the continuum electron needs to come back
 to revisit the parent ion in order to emit radiation, the neglect
 of the electron-ion interaction is rather questionable. Thus, various
 efforts have been made to improve upon the Lewenstein model, by
 including Coulomb distortion \cite{ivanov,kaminski96}, or by using Ehrenfest theorem
 \cite{gordon}. These improvements, however, still do not lead to satisfactory
agreement with exact calculations, nor do they reveal how the target
affects the HHG spectra.

  According to the three-step model, the last step of HHG
  is the recombination process. Taking this model seriously, one
  can ask if the HHG yield can be written as the product of a returning
  electron wave packet with the photo-recombination cross section, and if the
  electron wave packet is largely independent of the targets.
  These two assumptions are at the heart of the tomographic
  procedure of Itatani {\it et al.} \cite{itatani}.
  Their validity has been established for N$_2$
and O$_2$ molecules where the HHG spectra were calculated using the
Lewenstein model and the photo-recombination cross sections were
calculated using the plane-wave approximation (PWA) \cite{hoang}.
The independence of the returning wave-packet on the atomic rare gas
targets has recently been investigated by Levesque {\it et al.}
\cite{david}, where the dipole matrix elements (or recombination
cross sections) were also calculated within the PWA.

More recently, using HHG spectra obtained from numerical solutions
of the time-dependent Schr\"odinger equations (TDSE) for atoms in an
intense laser field, Morishita {\it et al.} \cite{toru} showed that
the ``exact'' TDSE results can be expressed as
\begin{equation}
S(\omega)=W(E) \frac {d\sigma(\omega)}{d\Omega_{{\bm k}}},
\end{equation}
where $d\sigma/d\Omega_{{\bm k}}$ is the ``exact''
photo-recombination (differential) cross section.   Here $W$
describes the flux of the returning electrons, which we will call a
``wave-packet". The electron energy $E$ is related to the emitted
photon energy $\omega$ by the relation $E=k^2/2=\omega-I_p$, with
$I_p$ being the ionization potential of the target (atomic units are
used throughout this paper, unless otherwise indicated). The method
for solving the TDSE and the calculation of $d\sigma/d\Omega_{{\bm
k}}$ are described later.

In this Letter, we will first show that if the HHG yield calculated
within the Lewenstein model is written as
 \begin{equation}
S^{SFA}(\omega)=W^{SFA}(E) \frac
{d\sigma^{PW}(\omega)}{d\Omega_{{\bm k}}},
\end{equation}
then the electron ``wave packet" $W^{SFA}$ will be largely identical
to $W^{TDSE}$ obtained from Eq.~(1) by solving the TDSE in the same
laser pulse. Here $d\sigma^{PW}/d\Omega_{{\bm k}}$ is the
recombination cross section calculated the plane wave approximation.
In the improved Lewenstein model, or the scattering wave based
strong-field approximation (SW-SFA), we propose that the HHG  be
calculated by
\begin{equation}
S^{SW-SFA}(\omega)=W^{SFA}(E) \frac
{d\sigma(\omega)}{d\Omega_{{\bm k}}}.
\end{equation}

\begin{figure}
\mbox{\rotatebox{0}{\myscalebox{
\includegraphics{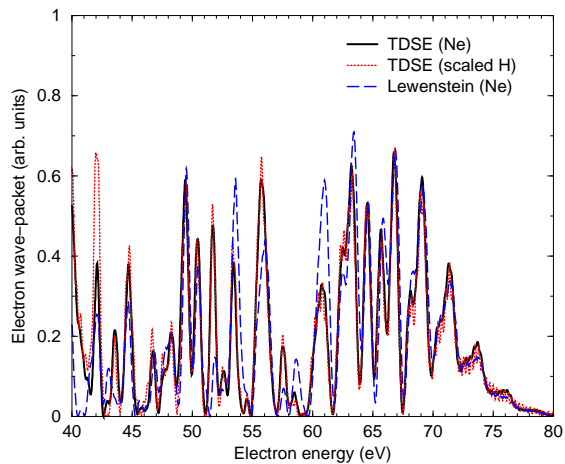}}}}
\caption{(Color online) Comparison of the electron ``wave
packets'' extracted from the exact HHG spectra of Ne, obtained by
solving the TDSE (solid black line), and from the Lewenstein model
(dashed blue line). Also shown is the TDSE result for a scaled H
(dotted red line). The laser pulse is with duration of 10.3 fs,
peak intensity of $2\times 10^{14}$ W/cm$^2$ and mean wavelength
of 1064 nm.} \label{fig1}
\end{figure}

In Fig.~1, we show the wave-packets extracted from the HHG spectra
of Ne (solid black line) and from scaled H (dotted red line) by
using Eq.~(1), where the ``exact'' HHG spectra are obtained from the
solution of the TDSE and the photo-recombination cross sections are
calculated with ``exact'' scattering wavefunctions. The effective
nuclear charge of the companion H is chosen such that it has the
same 1s binding energy as the Ne(2p) state. We use a laser pulse
with duration (FWHM) of 10.3 fs, peak intensity of $2\times 10^{14}$
W/cm$^2$ and mean wavelength of 1064 nm. Clearly, the two wave
packets agree very well. Additional evidences of such agreement can
be found in our recent work \cite{toru}, where Eq.~(1) was also
utilized to extract very accurate photo-recombination cross sections
for a variety of atomic targets. The extracted electron wave packet
$W^{SFA}$ is shown as the dashed blue line. This result, which has
been normalized to that of the TDSE near the cutoff, agrees
reasonably well with the TDSE result. We have also performed
calculations for different atoms and with different laser parameters
to confirm this conclusion.

\begin{figure}
\mbox{\rotatebox{0}{\myscaleboxa{
\includegraphics{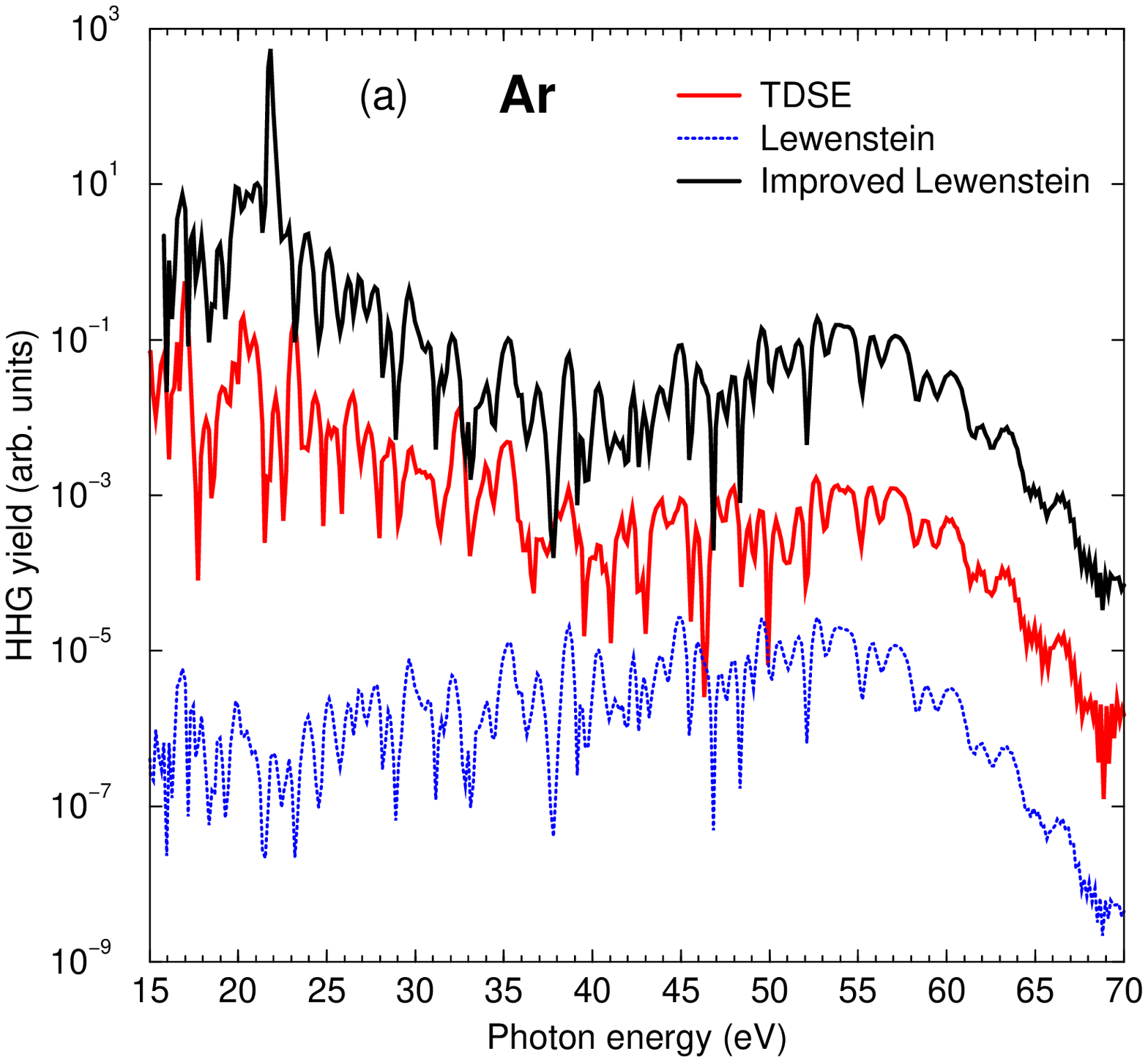}}}}
\mbox{\rotatebox{0}{\myscaleboxa{
\includegraphics{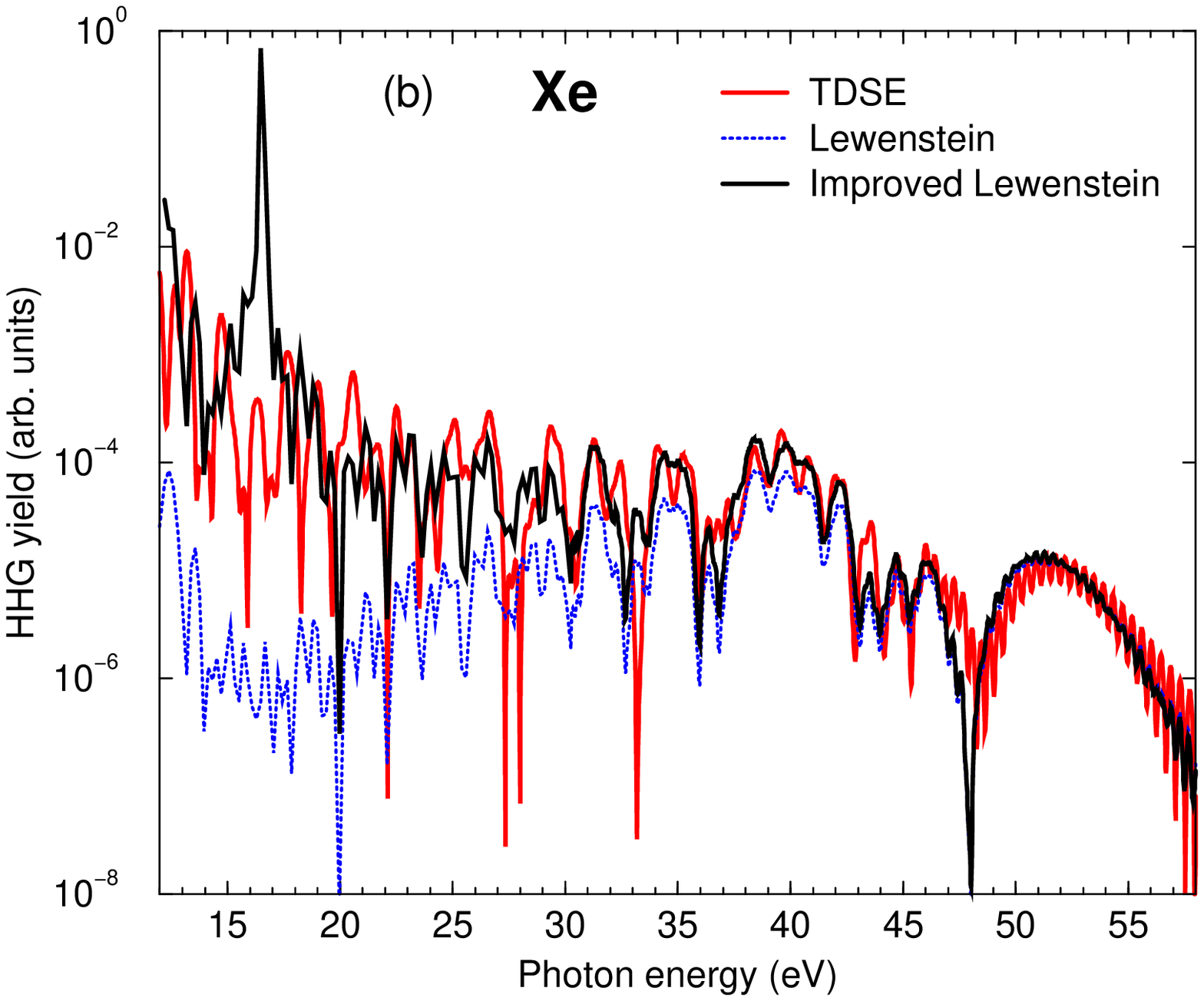}}}}
\mbox{\rotatebox{0}{\myscaleboxa{
\includegraphics{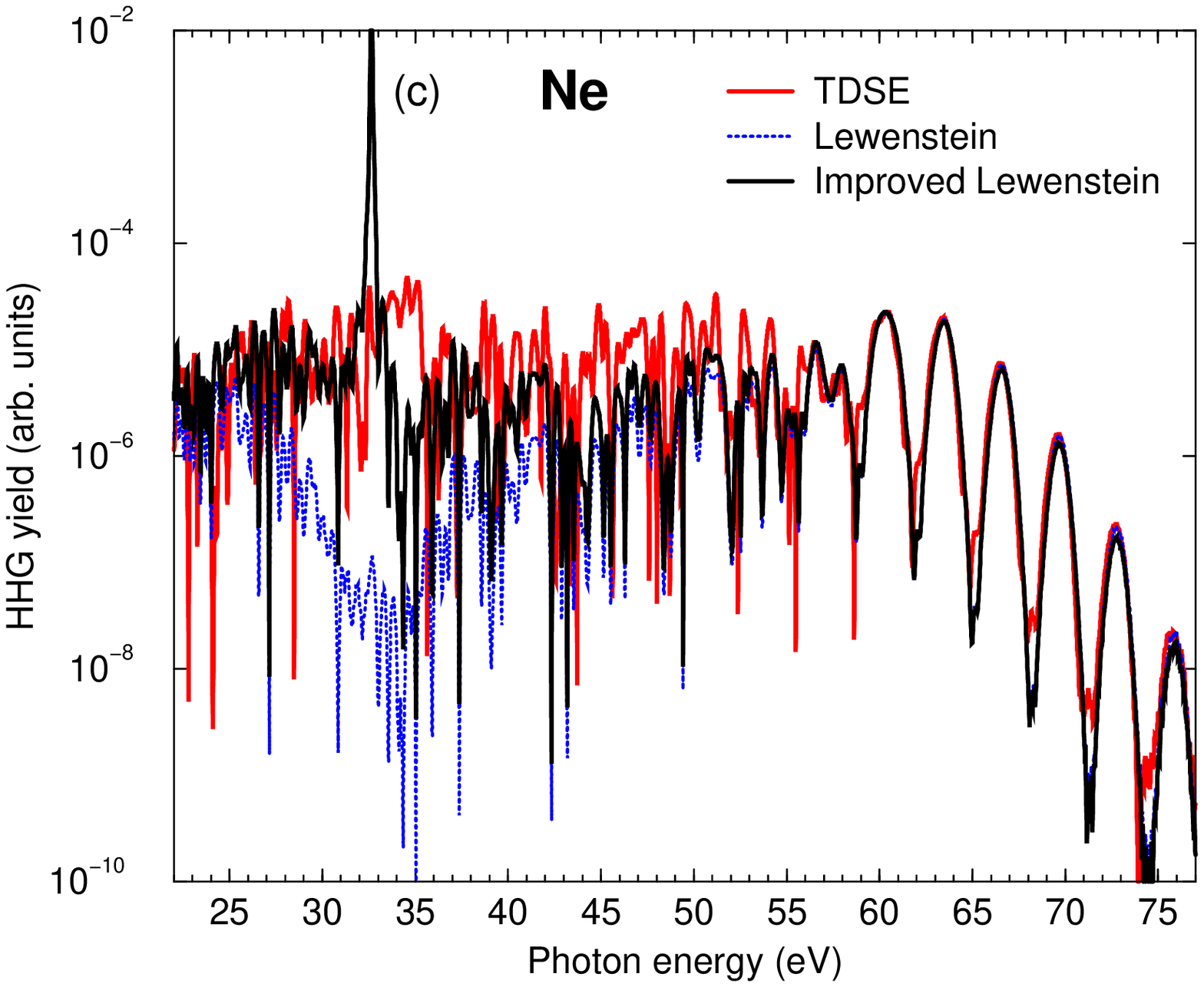}}}}
\caption{(Color online) Comparison of the HHG yields obtained from
numerical solution of the TDSE (solid red lines), the Lewenstein
model (dashed blue lines), and the improved Lewenstein model
(solid black lines) for Ar (a), Xe (b), and Ne (c).} \label{fig2}
\end{figure}

 Having established the good agreement between $W^{SFA}$ and
 $W^{TDSE}$, we now present the main results in Fig.~2.
 Here we show the HHG spectra calculated using the direct solution
of the TDSE (solid red lines), Lewenstein model (dotted blue
lines), together with the SW-SFA model (solid black lines) for Ar,
Xe, and Ne. For Ar and Ne, the laser pulse has peak intensity of
$2\times 10^{14}$ W/cm$^2$ and mean wavelength of 800 nm. The
laser duration (FWHM) is 10 fs for Ar and 20 fs for Ne. For Xe,
the corresponding parameters are $5\times 10^{13}$ W/cm$^2$,
1600nm, and 7.8 fs, respectively. The HHG yields for Ar are
shifted vertically in order to show their detailed structures. For
Ne and Xe, the SFA and SW-SFA results are normalized to the TDSE
results near the cutoff, i.e., close to $3.2U_p+I_p$, where $U_p$
is the ponderomotive energy.

First we compare the SFA and the TDSE results.  Clearly,  the
Lewenstein model agrees reasonably well only in some range near
the HHG cutoff for all three cases. For Ar, they agree well for a
broad range from 70 eV down to 40 eV. Near 40 eV, the slope of the
TDSE curve changes and the two results start to deviate. The
discrepancy reaches more than two orders of magnitude near the
ionization threshold (15.76 eV). For Xe, the Lewenstein model
underestimates the HHG yield by up to one or two orders of
magnitude, begining just below the cutoff. The inaccuracy of the
SFA for lower harmonics  is usually interpreted as due to the
neglect of the interaction between the target ion and the
continuum electron in the three-step model. While this
interpretation is qualitatively correct, it does not explain why
in some cases the agreement is better. For example, see Fig.~2(c)
for Ne. Here the HHG yield from the Lewenstein model disagrees
with the TDSE results only in the range from 28 eV to 45 eV.

We next examine the prediction of the SW-SFA model. From Fig.~2 it
is clear that it gives much better agreement with the TDSE results
for all three atoms shown.  In some cases, the discrepancy has been
reduced from two orders of magnitude in the SFA  to about a factor
of two in the SW-SFA. Since the recombination cross sections
calculated using ``exact'' scattering wavefunctions are independent
of the electromagnetic gauges used, unlike the SFA, the SW-SFA model
has no gauge-dependence issues.  There are some ``unphysical'' sharp
spikes in Fig.~2, near 22 eV in Ar,  17 eV in Xe, and 32 eV in Ne.
They are due to the ``Cooper minima'' where the photo-recombination
cross sections $d\sigma^{PW}/d\Omega_{{\bm k}}$ vanish (see Fig.~3).

\begin{figure}
\mbox{\rotatebox{0}{\myscalebox{
\includegraphics{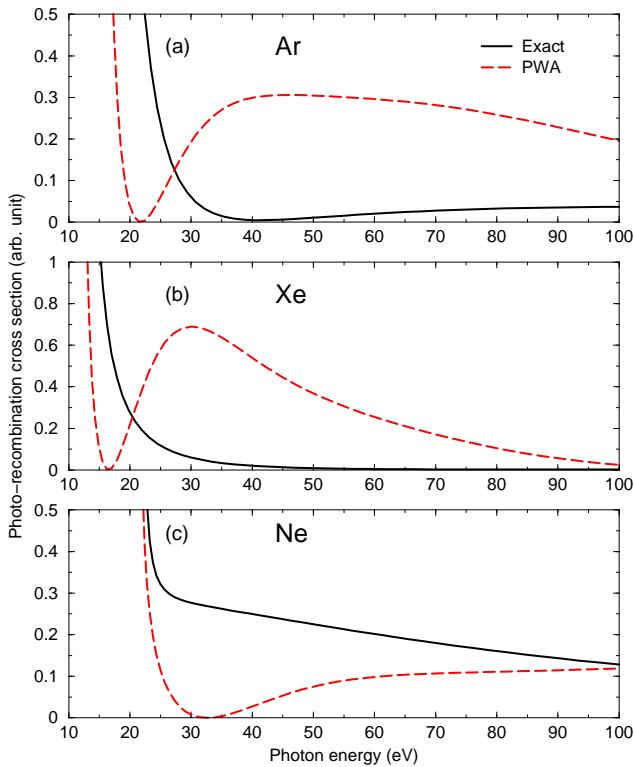}}}}
\caption{(Color online) Photo-recombination cross sections of Ar
(a), Xe (b) and Ne (c), obtained by using exact scattering
wavefunctions (solid curves) and within the plane-wave
approximation (dashed curves).} \label{fig3}
\end{figure}

In order to understand the nature of the improvement in the SW-SFA,
we now analyze the photo-recombination cross sections. In Figs.~3
(a), (b), and (c) we compare the exact results,
$d\sigma/d\Omega_{{\bm k}}$, with $d\sigma^{PW}/d\Omega_{{\bm k}}$
for Ar, Xe, and Ne, respectively. The figures show that, by
approximating scattering waves by plane waves, the calculated dipole
matrix elements are quite inadequate for electron energies below 100
eV. In particular, in Ne there is a spurious ``Cooper minimum'' near
32 eV if the plane wave is used. This minimum is totally absent for
the whole energy range in the exact result. The Cooper minima are
known to be related to the zeros in the dipole matrix elements and
they are absent for Ne since the 2p wavefunction is nodeless
\cite{starace}. We note, from Fig.~3(b), the position of the Cooper
minimum for Ar predicted by using the PWA is shifted lower by about
20 eV as compared to the exact result. For the Xe case, the shift is
even larger (the ``exact'' Cooper minimum, which is near 80 eV, is
not clearly seen in Fig.~3(b)).

 The theoretical method for the numerical solution of the TDSE
 has been described previously \cite{chen,toru1}.
 The electric field of the laser pulse is written
 in the form  $E(t)=E_0a(t)\cos(\omega t+\phi)$,
with the envelope given by $a(t)=\cos^2(\pi t/\tau)$, where $\tau$
is 2.75 times the FWHM of the laser pulse. All the atoms are
approximated by a one-electron model potential
 $V(r)$ which has the form of $V(r)=V_s(r)-1/r$.
 The parameters in the model potential are chosen such that the
binding energies of the ground state and the first few excited
states are close to the experimental values. We then propagate the
initial ground state in time using the split operator method. The
laser parameters were chosen such that the depletion effect is
small. Typically the ground state survival probability is more than
about 95\%.

Photo-recombination cross section is determined as
\begin{equation}
\frac{d\sigma(\omega)}{d\Omega_{{\bm k}}}=\frac{\omega^3}{2\pi
k}\left|\langle\Psi^+_{{\bm k}}|z|\Phi_0\rangle\right|^2,
\end{equation}
where $|\Psi^+\rangle$ and $|\Phi_0\rangle$ are the scattering
wavefunction and the ground state wavefuntion, respectively, and
$\Omega_{{\bm k}}\equiv \{\theta_{{\bm k}},\phi_{{\bm k}}\}$ is the
direction of ${\bm k}$. The returning electron that contributes most
to the HHG is the one that propagates along the laser polarization
direction. Therefore, the relevant cross section is for ${\bm k}$
that is parallel to $z$-axis (the laser polarization axis), that is,
$\theta_{{\bm k}}=0$ or $\pi$.

For each model potential $V(r)$, continuum scattering wavefunctions
are calculated numerically, using the partial-wave expansion
\begin{equation}
\Psi^+_{{\bm k}}({\bm r})=\sum_{l=0}^{\infty}\imath^l
e^{\imath\delta_l}\frac{R_{kl}(r)}{k}\sum_{m=-l}^l
Y_{lm}^*(\theta_{{\bm k}},\phi_{{\bm k}})Y_{lm}(\theta,\phi)
\end{equation}
where $\delta_l(k)$ is the partial-wave phase shift. With this
scattering wave function, ``exact'' photo-recombination cross
section is then obtained by using Eq.~4. Note that for atoms which
have  p$_0$ ground states, only $l=0,2$ and $m=0$ will contribute
to the dipole matrix element. It is interesting to note that for
all the systems considered here, the shape of
$d\sigma/d\Omega_{{\bm k}}$ as functions of photon energy are
quite close to the total (integrated) cross sections, obtained by
averaging over $m=-1,0,1$ initial states. In the PWA, a plane-wave
$(2\pi)^{-3/2}\exp(\imath{\bm k} {\bm r})$ is used to obtain
$d\sigma^{PW}/d\Omega_{{\bm k}}$.

\begin{figure}
\mbox{\rotatebox{0}{\myscaleboxb{
\includegraphics{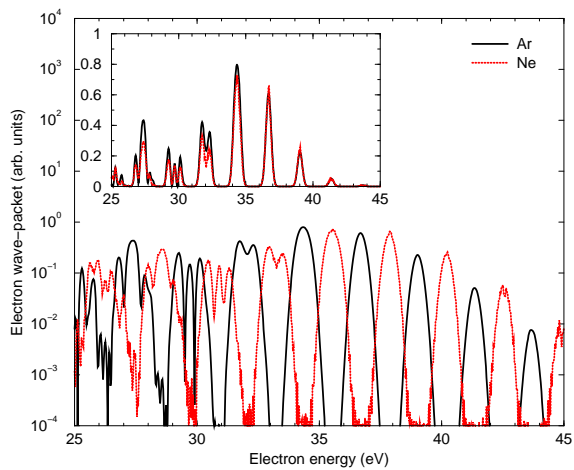}}}}
\caption{(Color online) The electron wave-packets, extracted from
the HHG spectra for Ne and Ar under the same laser field. The
inset shows the wave-packets in the linear scale with the Ne data
shifted horizontally by -1.2 eV. The HHG are obtained within the
Lewenstein model with 1064 nm laser, peak intensity of $2\times
10^{14}$ W/cm$^2$, duration of 50 fs.} \label{fig4}
\end{figure}

So far we have extracted photo-recombination cross sections from Ar,
Xe, and Ne by using companion atoms with nearly identical ionization
potentials, as compared to the atoms under consideration
\cite{toru}. This is used to greatly simplify the analysis. In the
following, we will show that this requirement can be relaxed. As the
wave packets obtained within the Lewenstein model agree reasonably
well with the TDSE results, for our purpose we only use the
Lewenstein model in the subsequent analysis. In Fig.~4, we show the
comparison of the wave packet from Ne ($I_p=21.56$ eV) and Ar
($I_p=15.76$ eV) in the same 1064 nm  laser pulse with peak
intensity of $2\times 10^{14}$ W/cm$^2$, duration (FWHM) of 50 fs.
Clearly, the two wave packets lie nicely within a common envelope.
This can be seen even more clearly in the inset, where the data are
plotted in linear scale with the Ne data shifted horizontally by
-1.2 eV. We note that the small details below 33 eV also agree well.
This conclusion is also confirmed by our calculations using the TDSE
with different laser parameters and with other atoms. This supports
that the independence of the wave packet on the target structure can
also be extended to systems with different ionization potentials.

 In this Letter, we have shown that the improved Lewenstein model is capable of
describing HHG spectra close to the accuracy that can only be
achieved by solving the TDSE directly. Extending this model to
molecules, we expect that accurate HHG spectra from molecular
targets can be calculated. For molecules, accurate solutions of the
TDSE are very time-consuming, if not impossible. Using the
Lewenstein model, the HHG as well as the dipole matrix elements can
be easily calculated \cite{zhou,atle} to extract the wave packet.
Similarly, theoretical tools for the calculation of molecular
photo-recombination (or photoionization) cross sections are quite
mature, and general computer codes are available even for large
molecules \cite{tonzani}. Thus it is expected that HHG spectra
generated by molecules can be calculated using the SW-SFA model
readily. Finally, we mention that the present analysis has not yet
addressed the phases of the harmonics. Further studies would be
needed to disentangle the phases due to the laser and the target
separately. This step is needed in order to incorporate propagation
effect of the medium on the generated high harmonics.

This work was supported in part by the Chemical Sciences,
Geosciences and Biosciences Division, Office of Basic Energy
Sciences, Office of Science, U. S. Department of Energy. TM is
also supported by a Grant-in-Aid for Scientific Research (C) from
the Ministry of Education, Culture, Sports, Science and
Technology, Japan,  by the 21st Century COE program on ``Coherent
Optical Science'',  and by a Japanese Society for the Promotion of
Science (JSPS) Bilateral joint program between US and Japan.

\end{document}